\documentclass[twocolumn,preprintnumbers,prd,superscriptaddress,nofootinbib,aps,10pt]{revtex4-1}

\usepackage{subfigure}
\usepackage{comment}
\usepackage[colorlinks=true,
            urlcolor=blue,
            anchorcolor=blue,
            citecolor=blue,
            filecolor=blue,
            linkcolor=blue,
            menucolor=blue,
            linktocpage=true,
            pdfproducer=medialab,
            pdfa=true]{hyperref}
\usepackage{amssymb,amsmath,amsfonts,multirow} 
\usepackage{graphicx}
\usepackage{bm}
\usepackage{setspace}
\usepackage[T1]{fontenc}
\usepackage[compat=1.1.0]{tikz-feynman}
\begin{document}

\title{Probing Solar Symmetrons with Direct Detection}

\author{Hannah Banks}
\email{hannah.banks@nyu.edu}
\affiliation{DAMTP, University of Cambridge, Wilberforce Road, Cambridge, CB3 0WA, UK \looseness=-1}
\affiliation{Center for Cosmology and Particle Physics, Department of Physics, New York University, New York, NY 10003, USA
\looseness=-1}

\author{Anne-Christine Davis}
\email{a.c.davis@damtp.cam.ac.uk}
\affiliation{DAMTP, University of Cambridge, Wilberforce Road, Cambridge, CB3 0WA, UK \looseness=-1}

\author{Luca Visinelli}
\email{lvisinelli@unisa.it}
\affiliation{Dipartimento di Fisica ``E.R.\ Caianiello'', Universit\`a degli Studi di Salerno,\\ Via Giovanni Paolo II, 132 - 84084 Fisciano (SA), Italy \looseness=-1}
\affiliation{Istituto Nazionale di Fisica Nucleare - Gruppo Collegato di Salerno - Sezione di Napoli,\\ Via Giovanni Paolo II, 132 - 84084 Fisciano (SA), Italy \looseness=-1}

\begin{abstract}
We provide the first investigation of the solar production of symmetrons, a well-motivated class of screened scalar fields with density dependent couplings to the Standard Model, and their subsequent absorption in underground direct detection experiments. We compute the flux of symmetrons produced through photon conversion in the magnetic field of the solar tachocline, and constrain the resulting luminosity to not exceed 3\% of the observed solar output. Even under the conservative assumption that production occurs only in the tachocline, this criterion yields robust astrophysical bounds on previously uncharted regions of symmetron parameter space, and predicts a keV-scale symmetron spectrum at Earth. We then derive the corresponding absorption signal in liquid xenon detectors, where symmetrons can interact with electrons through both conformal and disformal couplings. Using binned data from XENONnT, we obtain new direct-detection limits that are complementary to the solar luminosity constraint, further tightening the viable symmetron parameter space. Our results demonstrate that the Sun provides a testable, previously unexploited, source of symmetrons, and highlight that the interplay of astrophysical and laboratory searches offers a powerful strategy for probing screened scalar theories.
\end{abstract}

\maketitle

\section{Introduction}
\label{sec:introduction}

A compelling portfolio of evidence, encompassing observations that span from the galactic to the Hubble scale, testifies the existence of two additional components to the Universe: dark matter (DM) and dark energy (DE). The former proves necessary to explain a wide range of distinct phenomena from galactic rotation curves~\cite{Rubin:1970zza} to the cosmic microwave background (CMB) by way of its gravitational interactions with normal matter, whereas the latter drives the accelerated expansion of the Universe~\cite{Riess:1998cb,Perlmutter:1998np}. Despite decades of sustained theoretical and experimental effort, the nature of these entities, which vastly dominate over the visible Universe in terms of energy budget, remains largely unknown.    

The simplest explanation for DE is a cosmological constant resulting from the collective motion of the zero point energy of quantum fields. The naive prediction from quantum field theory exceeds the measured value by several orders of magnitude however, leading to alternative explanations being sought~\cite{Weinberg:1988cp}. Moreover, recent results from the Dark Energy Spectroscopic Instrument (DESI) experiment~\cite{DESI:2024mwx} appear to indicate that the equation of state of DE may evolve over time. If confirmed, this result would rule out the cosmological constant as the source of DE (see however Ref.~\cite{Efstathiou:2024xcq}). Although generally considered to be particulate in nature, DM remains similarly elusive at the microscopic level, having evaded detection in every direct and indirect search to-date. A cornucopia of possible candidates ranging from wave-like sub-eV bosons to $\sim 10^{19}$\,GeV mass primordial black holes remain phenomenologically viable. Taking inspiration from the rich microscopic offerings of the visible sector, there is a growing consensus that DM may just be one part of an extended suite of new states populating a range of scales. In many models these additional states provide a portal to the visible Universe by mediating DM-SM interactions, and are thus a phenomenologically relevant search target which may be able to shed new light on the dark side of the Universe.     

Light scalar fields are ubiquitous in theoretical constructions purporting to explain various aspects of the dark sector. They arise in many DM models, both as DM candidates and portal mediators, and have also been proposed to source the accelerated expansion of the late Universe~\cite{Ratra:1987rm, Zlatev:1998tr, Copeland:2006wr}. Any new light scalar boson coupling to SM matter fields generically gives rise to a  long range fifth force. Such phenomena are tightly constrained by a variety of laboratory and astrophysical tests of gravity, including bounds on post-Newtonian parameters and violations of the equivalence principle. To evade these limits, which effectively force any scalar-matter  coupling to be far weaker than gravity~\cite{Will:2005va, Burrage:2017qrf}, a variety of different screening mechanisms have been proposed. These include the chameleon mechanism~\cite{Khoury:2003aq, Brax:2004qh}, the symmetron~\cite{Hinterbichler:2010es, Brax:2011pk}, the Damour–Polyakov mechanism~\cite{Damour:1994zq, Brax:2010gi, Burrage:2025grx}, and the Vainshtein mechanism~\cite{Vainshtein:1972sx}. In the chameleon model, the scalar mass increases with density, suppressing its range  in high-density environments. In contrast, the symmetron model is built around a $\mathbb{Z}_2$ reflection symmetry. The symmetron potential and coupling to the SM conspire in such a way that below a critical density, which depends on the model parameters, the symmetry is spontaneously broken and the symmetron field acquires a vacuum expectation value (VEV). The leading effective coupling of the symmetron to SM fields is proportional to its VEV. As a result, if the local matter density is sufficiently high so as to restore the underlying $\mathbb{Z}_2$ symmetry and cause the VEV to vanish, the symmetron effectively decouples from the SM. It is this environmental dependence of the dominant symmetron-SM coupling: present in sparse environments but switching off or `screened' in dense environments, which permits the theory to evade the aforementioned constraints from  fifth forces searches and tests of local gravity.  Both the chameleon and symmetron mechanisms have been extensively studied and constrained in various laboratory, astrophysical, and cosmological contexts to-date~\cite{Davis:2011qf, Brax:2011aw, Upadhye:2012rc, Burrage:2018zuj, Feleppa:2025clx, Bartnick:2025lbg, Gan:2025icr, Feleppa:2025vop}, although limits on the symmetron mechanism are generally less restrictive and have yet to exploit the symmetron coupling to photons.  

Analogous to axion-like particles, light scalar fields which possess a coupling to photons can be produced in the Sun~\cite{Brax:2010xq, Redondo:2013lna, Redondo:2013wwa, Vinyoles:2015aba}. Experiments monitoring properties of the Sun and its output thus offer a novel opportunity to constrain these theories. Although originally designed with $\sim$ GeV mass WIMP-like DM particles in mind, underground direct detection experiments are also sensitive to scalar (and axionic) particles produced in the Sun, upon their arrival at the Earth. Indeed, in Ref.~\cite{Vagnozzi:2021quy} solar chameleons were identified to yield a possible explanation of the XENON1T anomaly~\cite{XENON:2020rca}. 

The study of the production of screened scalar fields in the Sun has so far been restricted to chameleons. Initial studies only considered production by photon conversion in the magnetic field of the solar tachocline, assuming that the chameleon mass became too large in the dense solar core for production there to contribute significantly to the overall flux. The expected emission spectrum from the tachocline has been searched for with the CERN Axion Solar Telescope (CAST) experiment and  constraints placed on the chameleon-photon coupling~\cite{Brax:2011wp}. Given the ultrahigh temperature in the solar core however, it was later realized in Ref.~\cite{OShea:2024jjw} that chameleon production in this region is not only non-negligible, but owing to the existence of additional production channels which become relevant at higher densities, can dominate over that mediated by the bulk magnetic field in the tachocline. As described in that work, incorporating production beyond the tachocline is more computationally intense given  the need to model the solar interior and magnetic field over an extended region. Updated computations of the solar chameleon flux including these additional channels, and its subsequent absorption in direct detection experiments, have since provided a crucial source of information on chameleon models, yielding novel constraints on their associated parameter space and complementing laboratory investigations which are typically insensitive to the chameleon-photon coupling~\cite{OShea:2024jjw,Yuan:2025twx}. 
In contrast, neither the production of symmetrons in the Sun, nor their potential interaction with underground direct detection experiments has been explored in literature at present. The Sun thus stands as a rich and currently unexploited test-bed to probe symmetron models.
Since the dominant effective interactions of the symmetron with the SM vanish upon restoration of the underlying $\mathbb{Z}_2$ symmetry, production only occurs in environments for which the ambient density is below the critical density of the theory. The viability of symmetron production in the tachocline, and its potential penetration further into the solar interior, is thus intrinsically parameter dependent. The symmetron differs from the chameleon in this respect, yielding a phenomenology which is correspondingly richer and more nuanced. 

In this work, we provide the first investigation of the production of symmetrons in the Sun, focusing exclusively on production in the magnetic field of the tachocline.  We derive new constraints on symmetron models both by considering the  percentage of the total solar luminosity emitted in symmetrons and the potential interaction of the emitted symmetron flux with underground direct detection experiments. We begin in Sec.~\ref{sec:methods} by discussing the relevant solar physics and how the Sun may be used as a laboratory to constrain or potentially detect scalar particles. We then introduce the symmetron model, outline the dominant  mechanism through which symmetron particles are produced in the tachocline  and compute the resulting luminosity as a function of the model parameters. In Sec.~\ref{sec:results} we present constraints on symmetron parameter space from consideration of the solar luminosity, before exploring direct detection prospects in the XENONnT~\cite{XENON:2022ltv} experiment. We discuss our results  in Sec.~\ref{sec:discussion} and comment on their relation to the wider landscape of symmetron studies. Our work constitutes the first examination of the symmetron-photon coupling, bridging a crucial gap in the literature on light scalar fields and offering unique access to large regions of well-motivated but currently uncharted parameter space. 


In this work we adopt the mostly-plus metric convention, work in natural units with $\hbar=c=1$ throughout and introduce the reduced Planck mass $M_{\rm Pl} = (8\pi G)^{-1/2}$.

\section{Methods}
\label{sec:methods}

\subsection{Solar physics}

The Sun has long served as a sensitive laboratory for testing new light particles. For axion-like (i.e. pseudo-scalar) particles, two main production mechanisms are relevant. The dominant channel is the Primakoff process, in which thermal photons in the solar plasma convert into axions in the Coulomb fields of charged particles~\cite{Raffelt:1990yz, Raffelt:1996wa, Redondo:2013wwa, DiLuzio:2020wdo}. The contribution from photon--axion conversion in solar magnetic fields adds up to the total flux~\cite{CAST:2004gzq, Jaeckel:2010ni}, although the comparatively poorly constrained magnetic field structure introduces large uncertainties to this component. The axion flux emitted from the Sun can be searched for through reconversion into photons in the magnetic fields of helioscopes such as the Tokyo Axion Helioscope~\cite{Inoue:2002qy}, CAST~\cite{CAST:2007jps, CAST:2015qbl, CAST:2015npk, CAST:2017uph, CAST:2018bce}, and in the proposed next-generation International Axion Observatory (IAXO) and BabyIAXO experiments~\cite{IAXO:2019mpb, IAXO:2025ltd}. Alternative approaches include the direct detection of solar axions via the axio-electric effect in underground detectors such as XENONnT~\cite{XENON:2022ltv}, PandaX~\cite{PandaX:2022ood, PandaX:2024sds, PandaX:2024cic}, and LUX-ZEPLIN (LZ)~\cite{LZ:2023poo, LZ:2025igz}. Similar stellar energy-loss arguments can also be applied to other astrophysical environments, such as red giants and white dwarfs~\cite{Dominguez:1999gg, Viaux:2013lha, Straniero:2020iyi, Dennis:2023aam, Dominguez:2025bgg}. These strategies and corresponding results provide a template for exploring other light scalar particles, including those with screening mechanisms.

Light screened scalar fields which possess a coupling to photons are also generically produced in, and emitted from the Sun. Studies of solar chameleon production have provided a wealth of unique information on these models, complementing other laboratory probes which more typically constrain the chameleon coupling to SM matter fields (fermions) as opposed to photons. An analogous investigation of the symmetron remains absent from the literature at present. Whilst it was shown in Ref.~\cite{OShea:2024jjw} that, across parameter space, chameleon production is viable throughout the solar interior, including the dense core, the story for symmetrons is more subtle. The dominant coupling of the symmetron to the SM is proportional to its VEV, which is non-zero \textit{only} in regions for which the ambient density lies below the parameter-dependent critical density of the theory. In order for symmetron production from photon conversion in the magnetic field of the tachocline, it is thus  necessary for the critical density of the theory to exceed the tachocline density. This requirement imposes an inherent restriction on the symmetron parameter space that can be probed with solar studies. Within the parameter space amenable to solar investigations, whether  or not production then penetrates further into the solar interior is also a parameter dependent statement. Owing to the monotonic rise in solar density towards the core, symmetron production generically continues inward until the local density surpasses the critical density of the theory upon which the resulting symmetry restoration drives the symmetron VEV to zero, effectively decoupling the field from the SM and terminating production.   

Given both the additional challenges of modeling production over an extended region, and that there has yet to be any investigation of the symmetron-photon coupling in literature to date, we restrict our attention to production in the  tachocline in this work, neglecting production further into the solar interior. Whilst this is a reasonable approximation for some of the  parameter space which we will consider, should the critical density significantly exceed the tachocline density such that production mediated by the symmetron VEV remains viable substantially further into the solar interior, our computation of the emitted symmetron flux is likely to 
underestimate  the true production rate. In this scenario, which we will characterize quantitatively in Sec.~\ref{subsec:sp}, our bounds should be viewed as a conservative first approximation which would likely become more stringent with the inclusion of additional production beyond the tachocline in future analyses. Given the very different modus operandi of the symmetron and chameleon screening mechanisms, the relative contributions of different regions of the Sun to the total emitted scalar flux are likely to differ between the two theories and it is not immediate that, even for regions of symmetron parameter space for which production in the core is viable, that this will dominate over that from the tachocline as in the chameleon case. We leave a more comprehensive study of symmetron production across the entire solar interior, including conversion both in the bulk magnetic field and  also via  Primakoff processes mediated by the magnetic field of charged ions  which become important at high densities, to future work. 

The solar tachocline consists of a thin ring shell at radius $R_t \approx 0.7\,R_\odot$, with radial extent $\Delta r \simeq [0.01, 0.05]\,R_{\odot}$. This region marks the transition between the radiative (to the interior) and convective (to the exterior) zones in the Sun and is believed to play a central role in the solar dynamo. The magnetic field strength in the tachocline is expected to reach up to $B \simeq 30$\,T, although this value is model-dependent and not directly measured~\cite{Bahcall:1995bt, Christensen-Dalsgaard:1996hpz, Gough:1996am, 1998Natur.394..755G}. In the presence of this magnetic field, photons can convert to symmetrons. To extract values for the necessary properties of the tachocline beyond its magnetic field we rely on non-magnetic standard solar models,  a variety of which exist in the literature \cite{Grevesse:1998bj, Asplund:2009fu, Caffau:2010qc, Scott:2014lka, Scott:2014mka, vonSteiger:2016ghw, Vagnozzi:2016cmr, Vinyoles:2016djt, Vagnozzi:2016pjr, Asplund:2021ghw}. Although they do not include large-scale magnetic fields, these models offer  robust determinations of the Sun’s interior structure and thermodynamic properties. In this work we specifically use the AGSS09 model to obtain values for the temperature and density of the tachocline ~\cite{Asplund:2009fu}.


The emission of new particles from the Sun can only account for a fraction of the total solar luminosity, $L_\odot \approx 3.83 \times 10^{26}$\,W, without conflicting with helioseismological observables and solar neutrino fluxes~\cite{Raffelt:1996wa, Raffelt:2006cw, Gondolo:2008dd}. For instance, CAST analyses adopt an upper bound of $10\%$ of the solar luminosity for exotic energy losses in their chameleon searches~\cite{Brax:2011wp}.  Although state-of-the-art standard solar models can now reproduce the observed luminosity at the $\sim 0.1\%$ level, the tachocline parameters necessary to compute scalar production therein are known with considerably lower precision. For this reason, quoting a single, universal luminosity fraction at such high precision without explicit justification would overstate the robustness of the resulting constraints. In this study we conservatively assume that the emitted symmetron luminosity cannot exceed     3\% of the total solar luminosity, matching the threshold employed in solar chameleon studies in Ref.~\cite{OShea:2024jjw}.

We now review the theoretical setup of the symmetron model, highlighting the density--dependent screening mechanism and the couplings relevant to solar production and detection in underground experiments.

\subsection{The symmetron}

As discussed, in the symmetron model~\cite{Hinterbichler:2010es, Brax:2011pk}, the dominant effective coupling to the SM depends on the ambient density. In high density environments, the coupling vanishes, whilst in low-density regions where the underlying symmetry is broken, it is restored. The transition between these two regimes is governed by the critical density of the theory, $\rho_*$, which is set by the model parameters. This screening mechanism is distinct from that of the chameleon~\cite{Khoury:2003aq, Brax:2004qh}, where the couplings to the SM remain finite but the scalar mass instead increases with density, thereby suppressing the range of the induced fifth force in dense environments.

The action describing the symmetron field $\phi$ in a scalar-tensor framework, written in the Einstein frame with metric $g_{\mu \nu}$, is 
\begin{equation}
    \label{eq:action}
    \begin{split}
    S =& \int {\rm d}^4 x \sqrt{-g} \left( \frac{M_\mathrm{Pl}^2}{2} R - \frac{1}{2} (\partial \phi)^2 - V(\phi) \right) \\
    &+S_\mathrm{m}[ \tilde{g}_{\mu\nu}, \psi]~.
    \end{split}
\end{equation}
Here, $S_\mathrm{m} = \int {\rm d}^4 x \sqrt{-\tilde{g}} \hspace{0.5em}\mathcal{L}_{\rm m}( \tilde{g}_{\mu\nu}, \psi)~$ denotes the action for SM fermions (i.e. `matter fields'), generically denoted $\psi$, which are assumed to couple minimally to the Jordan-frame metric, related to the Einstein metric by a conformal transformation $\tilde{g}_{\mu \nu} = A^2(\phi) g_{\mu\nu}$. This term thus gives rise to conformal interactions between the symmetron and the SM fermions. The symmetron potential takes the symmetry-breaking quartic form, 
\begin{equation}
    V(\phi) = 
- \frac{1}{2}\mu^2  \phi^2 + \frac{1}{4} \lambda \phi^4\,,
\end{equation}
where $\mu$ is an energy scale corresponding to the bare symmetron mass, and $\lambda$ is a dimensionless coupling constant. 

Given the conformal coupling of the symmetron to SM fermions in $S_{\rm m}$, the Euler-Lagrange equations for the symmetron field are controlled by an effective potential that depends on the local matter density $\rho$
\begin{equation}
    V_{\textnormal{eff}}(\phi) = V(\phi) + \rho\,A(\phi)\,.
\end{equation}
The conformal function $A$ is conventionally parameterized in terms of the symmetron-matter coupling $\beta_m$ as
\begin{equation}
    \label{eq:mattercoupling}
    A(\phi) = 1 + \frac{\beta_m}{2M_{\rm Pl}^2}\phi^2\,.
\end{equation}
Substituting Eq.~\eqref{eq:mattercoupling} into the effective potential yields
\begin{equation}
    V_{\textnormal{eff}}(\phi) = \frac{1}{2}\left(\frac{\rho \beta_m}{M_{\rm Pl} ^2} - \mu^2 \right) \phi^2 + \frac{1}{4}\lambda \phi^4\,,
\end{equation}
whose minimum then depends on the ambient matter density. For densities exceeding the critical value,
\begin{equation}
    \label{eq:criticaldensity}
    \rho_* = \frac{\mu^2 M_{\rm Pl}^2}{\beta_m}\,,
\end{equation}
the $\mathbb{Z}_2$  symmetry remains intact and the symmetron VEV vanishes. On the other hand for densities $\rho < \rho_*$ the reflection symmetry is spontaneously broken and the scalar acquires a VEV: 
\begin{equation}
    \label{eq:phi0}
    \langle \phi \rangle \equiv \phi_0 =
    \begin{cases}
    
    0\,,\quad\hbox{$\rho > \rho_*$}\,, \\ \frac{\mu}{\sqrt{\lambda}}\sqrt{1 - \frac{\rho}{\rho_*}}\,,\quad\hbox{$\rho \leq \rho_*$}\,.\\
    
    \end{cases}
\end{equation}
Expanding the symmetron field around its VEV as $\phi=\phi_0+\delta\phi$ in Eq.~\eqref{eq:mattercoupling}, 
it is straightforward to see that the leading interaction of the symmetron with SM matter fields is in fact linear in the symmetron fluctuation,
with an effective coupling constant proportional to the background field value:
\begin{equation}
    \label{eq:Lagrangian_matter}
    \mathcal{L}_{\rm m} \supset \frac{\beta_m\,\rho\,\phi_0}{M_{\rm Pl}^2}\,\delta \phi\,.
\end{equation}

The environmental dependence of this coupling constitutes the basis of the symmetron screening mechanism. 

Under the usual rules of EFT a coupling to photons that respects the symmetries of the model should also be included in the action of Eq.~\eqref{eq:action}. For the symmetron, the appropriate  Lagrangian density is 
\begin{equation}
    \label{eq:LagrangianEM}
    \mathcal{L}_{\gamma} \supset \frac{\beta_\gamma \phi^2}{2 M_{\rm Pl}^2} F^{\mu \nu}F_{\mu \nu}\,~,
\end{equation} 
where $F^{\mu \nu}$ denotes the SM electromagnetic field strength tensor. 

Unlike the chameleon, which couples linearly to photons through operators of the form $\propto\phi\,F^{\mu\nu}F_{\mu\nu}$, 
we emphasize that the symmetron coupling is necessarily quadratic as a result of the defining $\mathbb{Z}_2$ symmetry of the theory. Expanding the symmetron field around its local background value as $\phi = \phi_0 + \delta\phi$ then  yields, 
\begin{equation}
    \label{eq:LagrangianEM2}
    \mathcal{L}_{\gamma} \sim \frac{\beta_\gamma}{M_{\rm Pl}^2}(\phi_0\delta \phi + \frac{1}{2}\delta \phi^2) F^{\mu \nu} F_{\mu \nu}\,,
\end{equation}
analogously to the coupling to SM fermions. 
The leading interaction relevant for symmetron production and detection arises from the term linear in the symmetron fluctuation. As a result, the effective symmetron-photon coupling is suppressed by a factor $\phi_0/M_{\rm Pl}$ relative to the corresponding chameleon-photon coupling. This mapping correctly captures the fact that symmetron–photon interactions remain present only in sufficiently low density environments where the symmetry is spontaneously broken and the symmetron VEV is finite. 

A dimension-six operator such as the symmetron-photon coupling in Eq.~\eqref{eq:LagrangianEM} could plausibly arise from a variety of UV mechanisms. For instance, if the symmetron couples to heavy charged fermions $X$ of mass $M_X$ through a Yukawa interaction $\mathcal{L} \supset y_{\phi X}\phi \bar X X$, with coupling $y_{\phi X}$, an effective Lagrangian term of the form
\begin{equation}
    \mathcal{L}_{\rm eff} \sim \frac{y_{\phi X}^2}{16\pi^2M_X^2}\phi^2F^{\mu \nu} F_{\mu \nu}\,,
\end{equation}
is generated when the heavy fermions are integrated out at one loop,  
analogous to the generation of the Higgs--photon coupling in the SM via top-quark or $W$-boson loops. In addition, such couplings arise naturally in the context of string theory and extra-dimensional models, where the four-dimensional effective action after compactification typically contains a gauge kinetic function $f(\phi)$ which multiplies the gauge field strength, as
\begin{equation}
    \mathcal{L}_{\rm eff} \supset -f(\phi)\, F^{\mu\nu} F_{\mu\nu}\,,
\end{equation}  
where $\phi$ here denotes  a modulus field, such as the dilaton or geometric moduli of the internal manifold~\cite{Kaplunovsky:1995jw}. Expanding $f(\phi)$ around its VEV generates effective operators of the form $\phi^2 F^2$, with a suppression scale set by the compactification or string scale.

While the symmetron model is conventionally formulated with a purely conformal coupling to the SM, as in Eq.~\eqref{eq:action}, more general interactions are allowed from an effective field theory perspective. In particular, when considering the  phenomenological implications of the symmetron model beyond solar production, it is useful to briefly discuss the most general scalar-matter couplings allowed by fundamental principles.

Bekenstein showed that the most general coupling of a scalar field to matter fields consistent with causality and the weak equivalence principle takes the form~\cite{Bekenstein:1992pj}
\begin{equation}
    \label{eq:frames}
    \tilde{g}_{\mu\nu} = A^2(\phi,X)g_{\mu\nu}^E + B^2(\phi, X) (\partial_\mu\phi)(\partial_\nu\phi)\,,
\end{equation}
where as before $g_{\mu\nu}$ is the Einstein-frame metric and $X \equiv (\partial_\mu \phi)^2$ includes derivative self-interactions. The first term corresponds to the conformal coupling, as used in the standard symmetron construction, while the second term represents a disformal coupling arising from gradients of the scalar field. The disformal interaction generates additional couplings with matter and radiation through terms of the form
\begin{equation}
    \mathcal{L}_{\rm dis} \supset \frac{\partial_\mu \phi \partial_\nu \phi}{M_e^4}T^{\mu\nu}\,,
\end{equation}
where $T^{\mu\nu}$ is the energy–momentum tensor and $M_e$ is an energy cutoff scale. Bounds on the size of the disformal photon coupling from stellar energy loss arguments mean that for solar production, the disformal channel is typically subdominant to the conformal one within the parameter space of interest, with negligible contribution to the overall emitted symmetron luminosity~\cite{Brax:2014vva}. In contrast, in terrestrial direct detection experiments, disformal couplings of the symmetron to SM fermion fields can provide the leading observable signature and play a central role in constraining the model~\cite{Vagnozzi:2021quy, Yuan:2025twx}. We will return to this point in Sec.~\ref{sec:results}. 

Finally, symmetron fluctuations about the background value $\phi_0$ acquire an effective density-dependent mass as a result of the conformal coupling to matter,
\begin{equation}
    \label{eq:mass}
    m_s^2 = 2 \mu^2 \left(1 - \frac{\rho}{\rho_*}\right)\,.
\end{equation}
This expression, which is valid in the symmetry-broken phase $\rho<\rho_*$,  assumes a maximum value of $m_s^2 \approx 2\mu^2$ in the low-density limit, and  tends to zero as the critical density is approached from below. This relation provides a key link between the conditions governing symmetron production in the Sun and the prospects for directly detecting the emitted flux in underground experiments  on Earth. 

\subsection{Solar production of symmetrons}
\label{subsec:sp}

We now turn to the calculation of the symmetron production rate in the solar plasma. As previously discussed, in this work we focus exclusively on production from photon conversion in the bulk magnetic field of the tachocline. This imposes a lower limit on the critical density that we can probe, which must exceed the tachocline density so that the theory lies in the broken phase and production can proceed via the finite symmetron VEV.  

For a given $\mu$, this requirement translates to an upper limit on $\beta_m$ of
\begin{equation}
\label{eq:con}
    \beta_m \leq \frac{\mu^2 M_{\rm Pl}^2}{\rho_t}
    \approx 7\times 10^{36}
    \left(\frac{\mu}{\rm eV}\right)^2 \,,
\end{equation}
where $\rho_t \approx 0.2\,{\rm g\,cm^{-3}}$ denotes the tachocline density~\cite{Asplund:2009fu}. This stands as a hard cutoff in the parameter space that can be probed through solar production. Whilst this condition delineates all  parameter space for which  production in the tachocline is viable, we stress that it is only upon saturating this inequality, i.e. when $\rho_* \sim \rho_t$, that  production is confined to the tachocline, since all regions to the interior necessarily lie in the symmetry restored phase of the theory. For all other choices of $(\beta_m, \mu)$ satisfying Eq.~\eqref{eq:con}, however, $\rho_*$ exceeds $\rho_t$, and production extends into the denser interior, stopping only when $\rho(r) = \rho_*$. Since we only compute production in the tachocline in this work, our treatment should therefore be viewed as a conservative estimate of the total symmetron luminosity for all but the maximum value of $\beta_m$ shown on our plots. As we will demonstrate, this approximation nonetheless permits us to probe large regions of currently unexplored parameter space, with the potential for further improvements, particularly at low $\beta_m$, should this additional production be included in future analyses.   

Symmetron production in the tachocline primarily proceeds via photon conversion in the large-scale solar magnetic field. Due to the relatively low density of the tachocline, this process outweighs Primakoff-like conversion in the microscopic  magnetic fields of charged particles, which we therefore neglect from our computation. This latter channel can become important in higher-density regions however and must be accounted for when extending the analysis to include production further into the solar interior.


To estimate the  rate of symmetron production in the tachocline magnetic field, we adopt a thermal field theory approach, following methods previously applied to axion production~\cite{Caputo:2020quz, Guarini:2020hps} and to chameleons~\cite{OShea:2024jjw}. In this framework, the relevant interaction arises from the conformal symmetron-photon coupling in Eq.~\eqref{eq:LagrangianEM}, which in the presence of an external magnetic field reduces to
\begin{equation}
    \mathcal{L} \supset \frac{\beta_\gamma\phi_0}{M^2_\mathrm{Pl}} \mathbf{B} \cdot (\nabla \phi \times \mathbf{A})\,,
    \label{eq:lagrangian_TFT}
\end{equation}
where $\mathbf{A}, \mathbf{B}$ refer to the electromagnetic vector potential and the corresponding magnetic field, respectively. We neglect contributions from disformal couplings which are subdominant in magnetic-field-induced production processes due to constraints on the size of these couplings from stellar energy-loss arguments~\cite{Vagnozzi:2021quy}.

Using the methods described in Appendix~\ref{sec:appendix}, the differential production rate per unit energy integrated over the tachocline volume is
\begin{equation}
    \frac{\mathrm{d}\dot N}{\mathrm{d}\omega} \!=\! \frac{2 \beta_\gamma^2\phi_0^2}{\pi M_\mathrm{Pl}^4}  \frac{\omega(\omega^2 - m_s^2)^{3/2}}{(m_\gamma^2 - m_s^2)^2 \!+\! (\omega \Gamma_\gamma)^2} \frac{\Gamma_\gamma}{e^{\omega/T} - 1}\,  R_t^2 B_\perp^2(R_t) \Delta r\,,
    \label{eq:dN_TFT2}
\end{equation}
where $m_\gamma$ is the effective mass of the photon in the solar plasma, the function $\Gamma_\gamma$ accounts for the production and absorption of photons in the Sun by various processes, and $m_s$ is defined in Eq.~\eqref{eq:mass}. In obtaining this result, we have treated the tachocline as a thin shell of width $\Delta r$, centered at radius $R_t = 0.7 R_{\odot}$, and assumed that the bulk magnetic field and density remain approximately constant over this region. The total luminosity emitted in symmetrons is then 
\begin{equation}
    L_s =\int_{0}^{\infty}{\rm d}\omega\,\omega\,\frac{\mathrm{d}\dot N}{\mathrm{d}\omega}\,.
\end{equation}

The average differential flux of symmetron particles reaching the Earth is
\begin{equation}
    \label{eq:spec}
    \frac{\rm{d}\Phi_E}{\rm{d}\omega} = \left(\frac{1}{4}\right)\,\frac{1}{4 \pi d_{\rm sun}^2}\,\frac{\mathrm{d}\dot N}{\mathrm{d}\omega}~,
\end{equation}
where $d_{\rm sun}$ denotes the Earth-Sun distance and the factor of 1/4 arises from the geometric projection of the emitted symmetron flux onto the Earth's surface.
We will henceforth refer to this quantity, which is necessary to calculate the rate of symmetron absorption in direct detection experiments, as the symmetron spectrum at the Earth. 

Given that the quartic symmetron self-coupling $\lambda$ only enters these expressions through the VEV as $\phi_0 \propto 1/\sqrt{\lambda}$,  solar production is only sensitive to the effective parameter combination $\beta_\gamma/\sqrt{\lambda}$ and cannot constrain either coupling in isolation. Accordingly, the bounds on symmetron parameter space obtained in Sec.~\ref{sec:results} of this work will be expressed in terms of $\beta_\gamma / \sqrt{\lambda}$.

\section{Results}
\label{sec:results}

\subsection{Bounds from solar luminosity}

The production of symmetrons in the solar tachocline introduces an additional energy-loss channel whose magnitude is constrained by the measured solar luminosity. Following the conservative approach of Ref.~\cite{OShea:2024jjw}, we require the total symmetron luminosity to remain below $0.03\,L_\odot$ to ensure consistency between precision helioseismological observables and their predictions from solar models.  

\begin{figure}
    \centering
    \includegraphics[width=\linewidth]{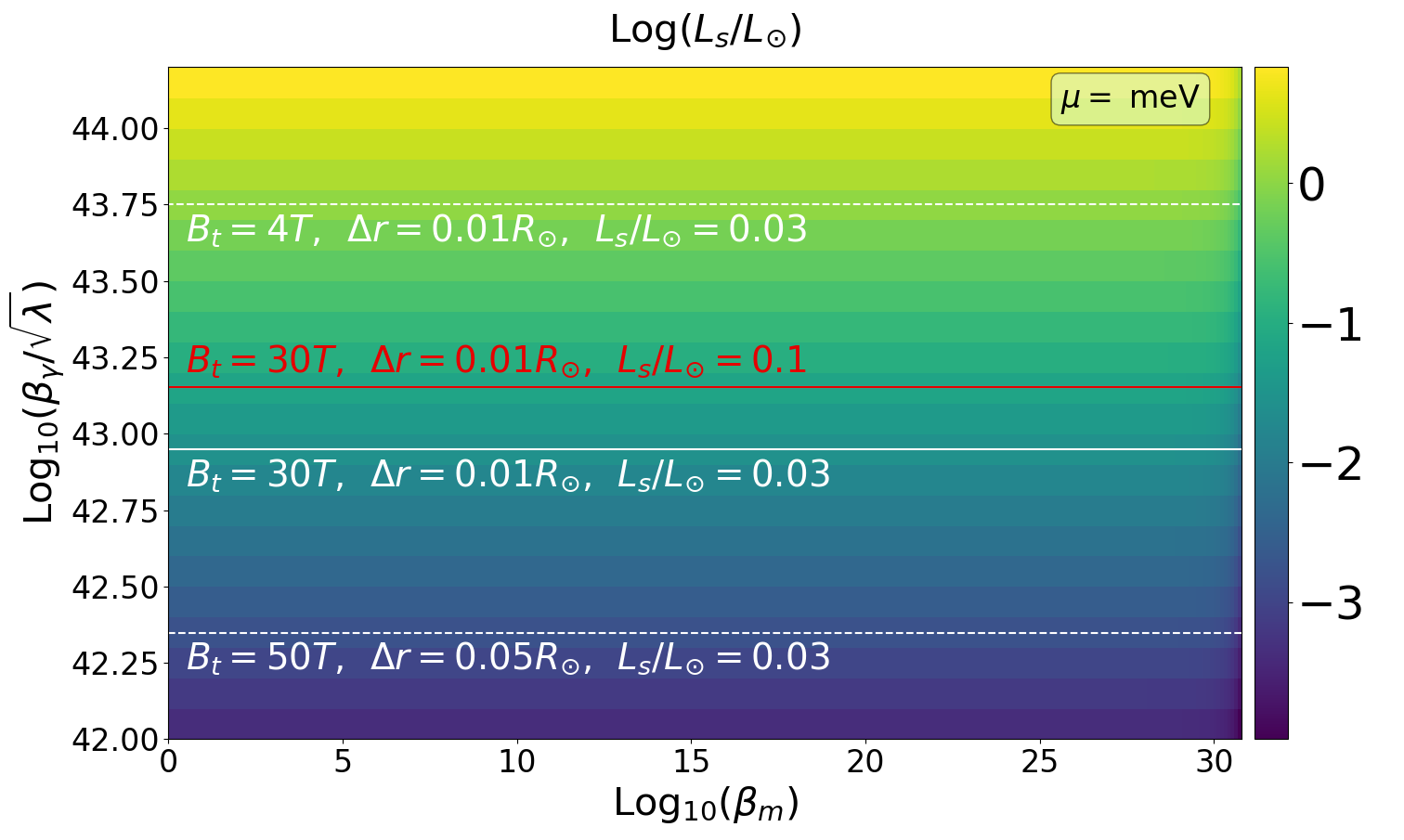}
    \caption{Bounds on symmetron parameter space for $\mu = 1$~meV from the requirement that the emitted symmetron luminosity does not exceed $3\%$ of $L_\odot$. The color-map displays the fraction of the total solar luminosity emitted in symmetrons at each point in parameter space, assuming the tachocline is of width $\Delta r = 0.01\,R_\odot$ and has a  magnetic field of $B_t = 30$\,T. The solid white line shows the $L_s/L_\odot = 0.03$ exclusion contour, with all parameter space above this line (i.e. with greater $\beta_\gamma/\sqrt{\lambda}$) deemed excluded. The solid red line plots the $L_s/ L_\odot = 0.1$ contour, to illustrate the effect of adopting a different exclusion threshold. The dashed white lines correspond to the envelope of the 3\% contour as $B_t \in [4,50]$\,T and $\Delta r \in [0.01,0.05]\,R_\odot$ are varied, capturing the spread in the bound induced by uncertainties on the tachocline parameters from solar modeling. }
    \label{fig:solar constraint}
\end{figure}

The color-map in Fig.~\ref{fig:solar constraint} shows the  fraction of the total solar luminosity emitted in symmetrons (i.e. $L_s/L_{\odot}$) across the $\beta_m$ -($\beta_\gamma/\sqrt{\lambda}$) plane for a fixed value of $\mu = 1$~meV, assuming a tachocline magnetic field $B_t = 30$\,T and width $\Delta r = 0.01\,R_\odot$. We plot the 3\%  contour for these tachocline parameters as a solid white line, and interpret all symmetron parameter space above this (i.e. with greater $\beta_\gamma/\sqrt{\lambda}$) as excluded. To characterize the impact of the uncertainty in the tachocline parameters on our exclusion bounds, we show with dashed white lines the envelope of the 3\% exclusion contours  when varying 
 $B_t \in [4,50]$\,T and $\Delta r \in [0.01,0.05]\,R_\odot$. These variations capture the range of solar-model predictions, which differ in the relative contribution of dynamo amplification and field configurations. As expected from the scaling of the symmetron production rate, larger field strengths and wider tachocline regions result in stronger constraints. 
 
 Given that various different luminosity fractions have been deployed as exclusion thresholds in studies of solar chameleons, for comparison purposes we also show the 10\% $L_s/L_{\odot}$ contour, which corresponds to the threshold used in the CAST chameleon search\footnote{Note that the line shown here denotes the constraint derived from demanding that no more than 10 \% of the solar luminosity is emitted in symmetrons and does not correspond to a bound from the CAST experiment, which would require modeling the back-conversion of this flux in the magnetic field of the CAST detector. We will extract a crude first estimate of the bound on the symmetron-photon coupling that could be obtained should  such an analysis be performed by rescaling the CAST chameleon limit in Section.~\ref{sec:discussion}, emphasizing that our estimate is subject to several caveats. }, as a solid red line. As shown, this contour falls within the uncertainty associated with modeling  the tachocline. As such, the specific choice of exclusion threshold adopted in this work is not expected to have a significant bearing on the conclusions drawn. Instead, the dominant uncertainty arises from the poorly constrained properties of the tachocline, emphasizing the importance of improving solar modeling in order to obtain more robust limits on new scalar fields. Finally, we stress that the choice of $\mu = 1$ meV in this figure is purely illustrative; luminosity constraints for a wider range of $\mu$ will be presented and compared with the corresponding limits from direct-detection experiments in Fig.~\ref{fig:directconstraints}. 

\begin{figure}
    \centering
    \includegraphics[width=\linewidth]{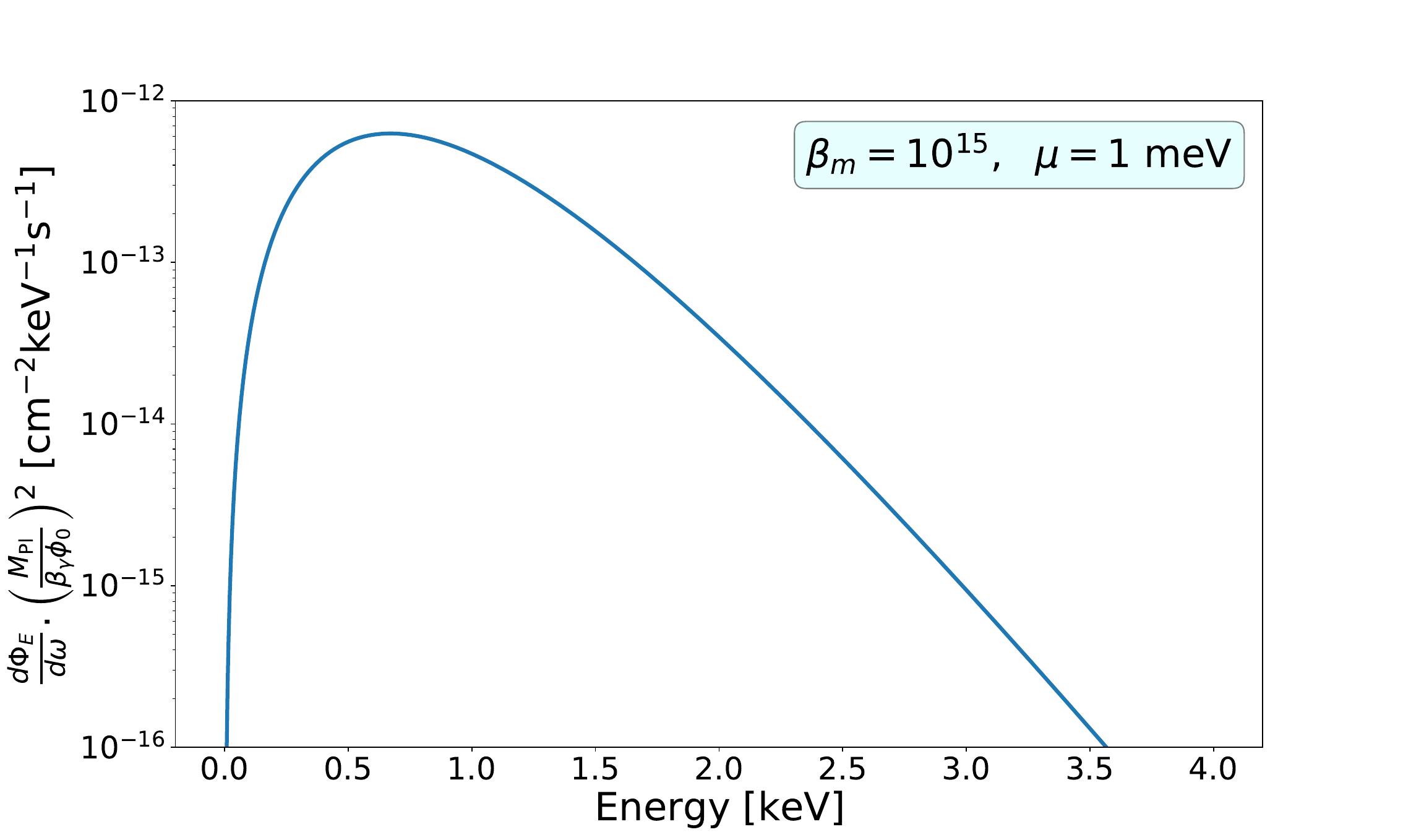}
    \caption{Predicted symmetron spectrum at Earth, normalized to remove the dependence on the couplings $\lambda$ and $\beta_{\gamma}$ for benchmark values $\beta_m = 10^{15}$ and $\mu = 1$ meV. The spectrum peaks in the keV range and provides the target flux for direct detection experiments.}
    \label{fig:symmetron spectrum}
\end{figure}

In Fig.~\ref{fig:symmetron spectrum} we display the solar symmetron flux reaching Earth, as defined in Eq.~\eqref{eq:spec}, after rescaling by the dimensionless factor $(M_{\rm Pl}/\beta_{\gamma} \phi_0)^2$. 
Since the physical spectrum scales as $(\beta_{\gamma} \phi_0/M_{\rm Pl})^2$,
this renormalization removes the dependency on 
$\beta_\gamma$ and $\lambda$ for convenience of visualization independent of these parameters.  As shown in the plot, the symmetron spectral profile, which we emphasize is independent of both $\beta_\gamma$ and $\lambda$, peaks in the keV range, similar to that of the solar axion and the chameleon. For illustration we have fixed $\beta_m = 10^{15}$ and $\mu = 1$ meV in this plot. We highlight that the magnitude of the physical spectrum (obtained by re-instating the $(\beta_{\gamma} \phi_0/M_{\rm Pl})^2$ factor which was removed in the normalization of this plot) directly tracks the effective symmetron-photon coupling in the tachocline, which is proportional to the local symmetron VEV.


\subsection{Bounds from symmetron detection}

The solar flux of symmetrons derived above provides a direct target for underground DM detectors. Once produced in the Sun, symmetrons reaching Earth can be absorbed in liquid xenon via interactions with electrons, generating keV-scale electronic recoils. Experiments such as XENONnT~\cite{XENON:2022ltv}, PandaX~\cite{PandaX:2022ood, PandaX:2024sds, PandaX:2024cic}, and LZ~\cite{LZ:2023poo, LZ:2025igz} are well suited to probe this signal, due to their large exposures and low energy thresholds. The absorption cross section consists of two distinct contributions, a conformal term proportional to the effective symmetron-electron coupling and a disformal term controlled by the energy scale $M_e$~\cite{Vagnozzi:2021quy}. Explicitly,
\begin{equation}
	\label{eq:symmetronelectric}
	\sigma_{\phi e} = \sigma_{\phi e, {\rm dis}} + \sigma_{\phi e, {\rm conf}}
	= \frac{m_e^2\omega^4}{8\pi^2\,M_e^8} + \frac{\beta_e^2\phi_0^2\omega^2}{2\pi\alpha M_{\rm Pl}^4}\,\sigma_{\rm photo} \,,
\end{equation}
where $\sigma_{\rm photo}$ denotes the xenon photoelectric cross section~\cite{Veigele:1973tza}, $m_e$ is the mass of the electron and, assuming the symmetron couples uniformly to the SM matter fields, $\beta_e = \beta_m$. The conformal term is suppressed relative to the corresponding chameleon-electron  coupling by the  factor $(\phi_0/M_{\rm Pl})^2$ and tends to zero as the local density approaches the critical density from below. In contrast the disformal term, which takes an identical form in chameleon theories, remains independent of $\phi_0$  and is therefore not subject to screening. Note that in order to obtain a symmetron flux at the Earth from solar production, we are necessarily probing symmetron parameter space for which the critical density exceeds the tachocline density and therefore also that of the detector. As such, the theory is in the broken phase and conformal interactions are unscreened. Since the effective conformal coupling is proportional to the local symmetron VEV however, the size of these interactions depends sensitively on the parameters $\mu$ and $\lambda$ through the combination $\mu/\sqrt{\lambda}$. The disformal interactions are independent of these parameters and can thus remain sizable even in regions of parameter space in which the conformal channel is suppressed due to small values of $\phi_0$. In this way direct detection can provide access to qualitatively different parts of the symmetron parameter space than can be reached with experiments relying on conformal interactions alone.

Given that the disformal term grows rapidly with energy as $\omega^4/M_e^8$, it typically  dominates the high-energy tail of the spectrum across the parameter space probed. At lower energies however, the relative importance of the conformal and disformal channels depends sensitively on the value of $M_e$, the inverse of which sets the size of disformal symmetron-electron coupling, relative to the effective conformal coupling $\sim \beta_m \phi_0$.  The difference between how the conformal and disformal couplings respond to the local environment is central to the phenomenology of the symmetron model and provides an additional handle with which to distinguish these fields from axion-like particles, whose absorption is entirely governed by $\sigma_{\rm photo}$~\cite{XENON:2022ltv}. 

The decomposition in Eq.~\eqref{eq:symmetronelectric} directly follows from the general metric coupling of Eq.~\eqref{eq:frames}. The conformal coupling generates operators proportional to $A(\phi) T$, where $T$ is the trace of the energy-momentum tensor.  As a result one generically expects a coupling to electrons of the form $\sim A(\phi) m_e \bar{e}{e}$, where  we have used the on-shell expression for the trace of the energy-momentum tensor. For consistency with the reflection symmetry of the theory, the minimal choice of the conformal factor $A$ consists of a quadratic in $\phi$ as per Eq.~\eqref{eq:mattercoupling}. As discussed previously, when expanded around the local background field value, the term linear in the symmetron fluctuation has an effective coupling constant of $\beta_m \phi_0/M_{\rm Pl}^2$, which then enters the cross-section quadratically, giving rise to the second term in Eq.~\eqref{eq:symmetronelectric}.\footnote{Note that for chameleons, the coupling to electrons is intrinsically linear arising from the choice $A(\phi) = (\beta_m/M_{\rm Pl})\phi$. The difference in the power of $M_{\rm Pl}$ relative to the symmetron case is necessary on dimensional grounds. As such the conformal cross-section for the symmetron in Eq.~\eqref{eq:symmetronelectric} is suppressed by a factor of $(\phi_0/M_{\rm Pl})^2$ relative to analogous cross-section for the chameleon.} By contrast, the minimal choice for $B(\phi)$ corresponds to a constant, generating a disformal interaction of the form $(\partial\phi)^2 T^{\mu\nu}$. Being a derivative interaction, this does not receive any additional suppression factors upon expanding about the VEV and thus leads directly to the first term in Eq.~\eqref{eq:symmetronelectric}. The differing structure of the disformal and conformal contributions has an important phenomenological consequence: only the conformal contribution allows for a  rescaling of the photoelectric cross section, whereas the disformal operator couples to the full stress--energy tensor and thus cannot be inferred from photoabsorption alone~\cite{Dimopoulos:1986mi, Dimopoulos:1986kc, Pospelov:2008jk}. For completeness we note that since  the leading order disformal contribution to Eq.~\eqref{eq:frames} is already quadratic in the scalar field, the minimal disformal symmetron-matter and chameleon-matter couplings (which both arise from the choice $B$ = constant) are identical.

The expected rate of symmetron absorption in the detector is obtained by folding the solar flux with the absorption cross section and the detector response,
\begin{equation}
    \label{eq:drdomegath}
    \left ( \frac{\mathrm{d}R}{\mathrm{d}\omega} \right )_{\rm th} = \epsilon(\omega)\,\int \frac{\mathrm{d}R_0(\omega_R)}{\mathrm{d}\omega_R}\,
    \Theta(\omega-\omega_R)\,\mathrm{d}\omega_R\,,
\end{equation}
where $\epsilon(\omega)$ is the detection efficiency specified by each experiment. For XENONnT the efficiency, corrected for event-selection cuts, is given in Fig.~1 of Ref.~\cite{XENON:2022ltv}. The unsmeared rate before folding in the detector response is obtained by combining the solar flux with the absorption cross section of Eq.~\eqref{eq:symmetronelectric},
\begin{equation}
    \label{eq:rawrate}
    \frac{\mathrm{d}R_0}{\mathrm{d}\omega} = \frac{1}{M_{\rm Xe}} \, \frac{\rm{d}\Phi_E}{\rm{d}\omega} \, \sigma_{\phi e}(\omega) \,,
\end{equation}
where $M_{\rm Xe}$ is the xenon target mass and 
the solar symmetron flux at the Earth is defined in Eq.~\eqref{eq:spec}.

As a representative case, we focus on the binned electron recoil data reported by the XENONnT collaboration~\cite{XENON:2022ltv}, which we take as the benchmark for current xenon-based direct-detection experiments. Given the comparable thresholds, backgrounds, and exposures achieved by PandaX and LZ, their sensitivities to solar-produced symmetrons are expected to be of the same order of magnitude as XENONnT. In practice, the most stringent limit arises from the lowest energy bin, whose large event statistics and high efficiency dominate the overall sensitivity once the published binning and detector response are taken into account. We present constraints in the parameter space spanned by $\beta_m$ and $\beta_\gamma/\sqrt{\lambda}$, evaluated for different choices of the disformal energy scale $M_e$ and the fundamental scale $\mu$. Exploring these scenarios reflects the theoretical uncertainty in the values of these parameters and thus captures the range of possible phenomenological outcomes.

\begin{figure*}[htb]
    \centering
    \subfigure[\,$\mu = 0.1$\,meV]{
        \includegraphics[width=0.45\linewidth]{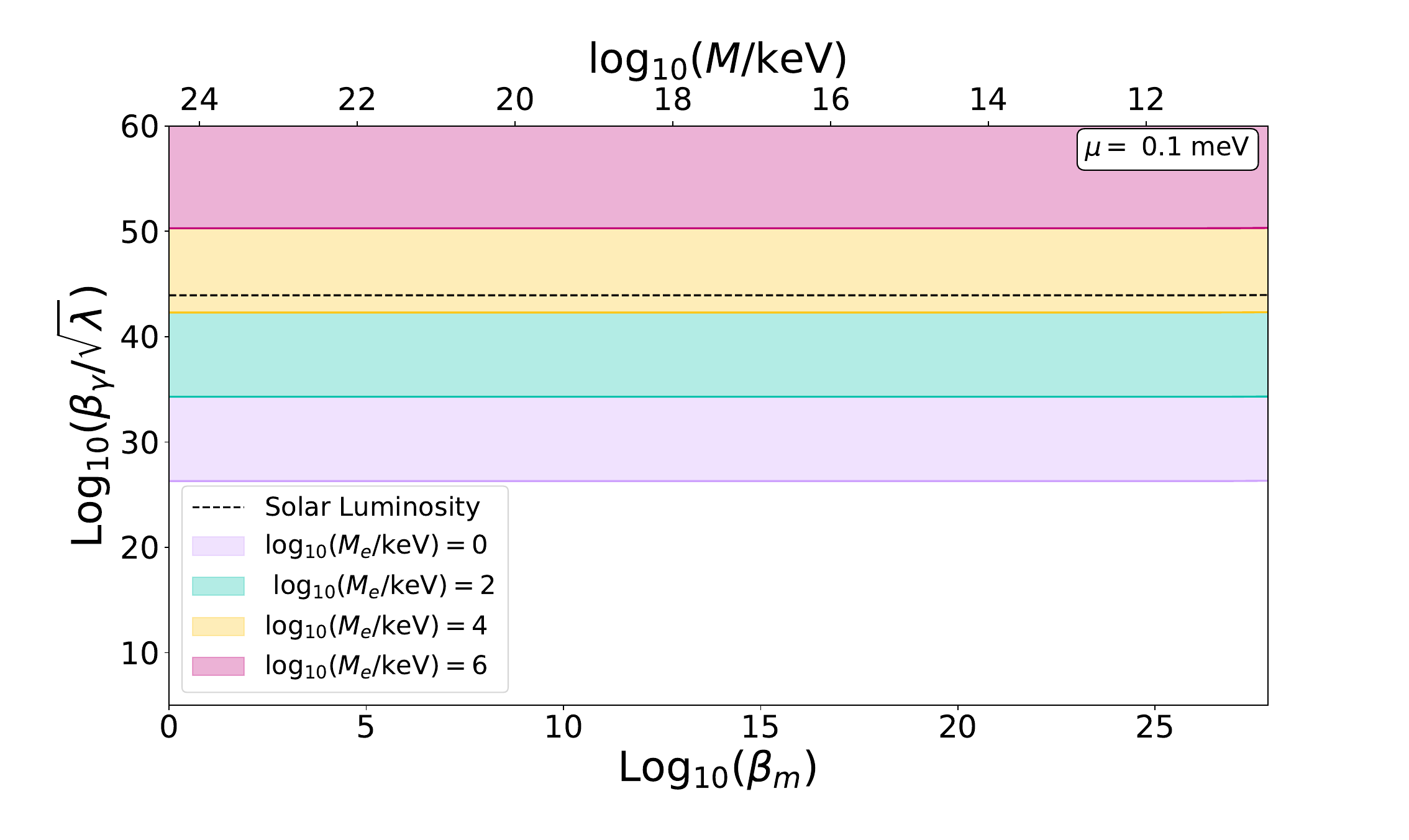}
        \label{fig:0.1meV}
    }
    \subfigure[\,$\mu = 1$\,meV]{
        \includegraphics[width=0.45\linewidth]{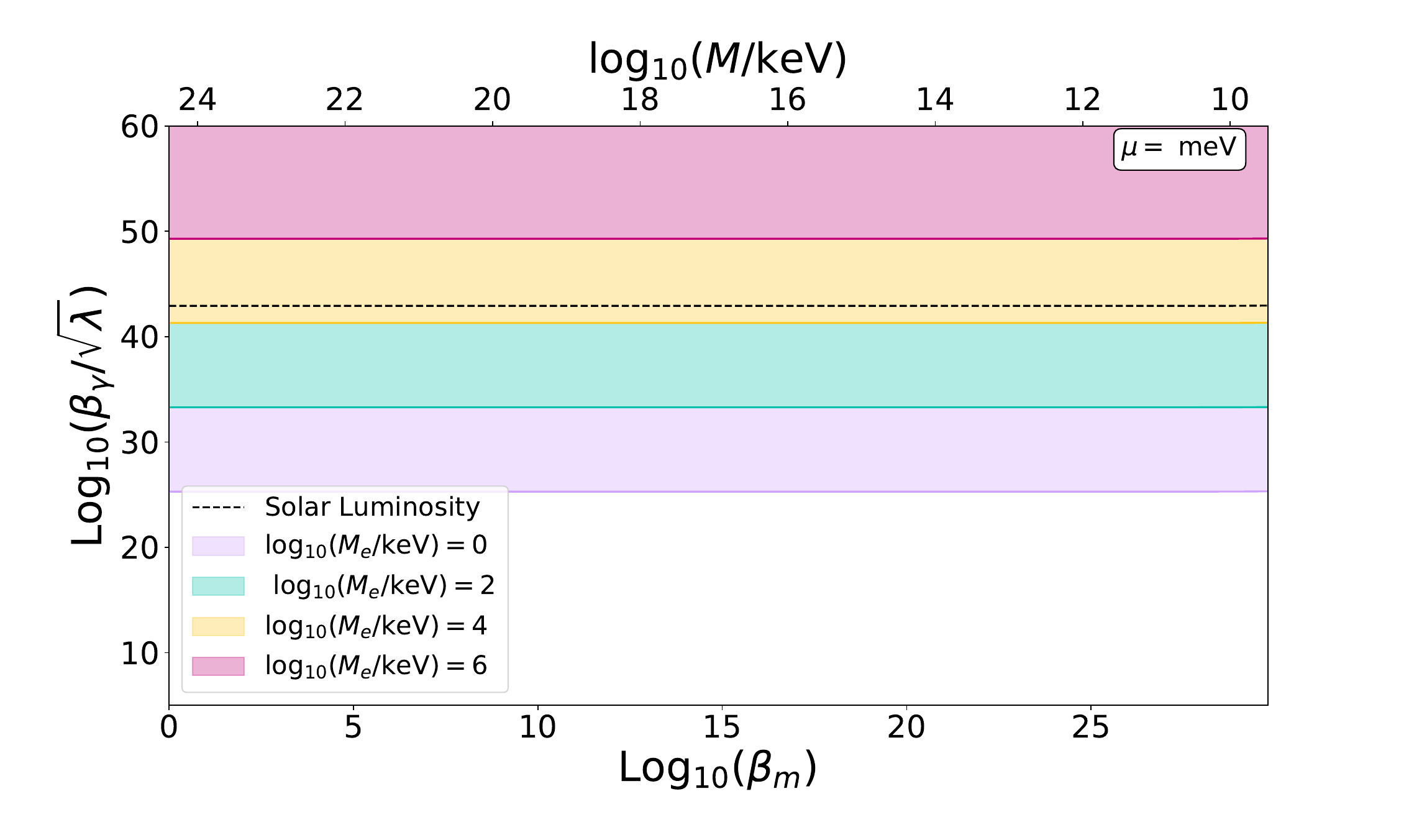}
        \label{fig:meV}
    }
    \subfigure[\,$\mu = 1$\,eV]{
        \includegraphics[width=0.45\linewidth]{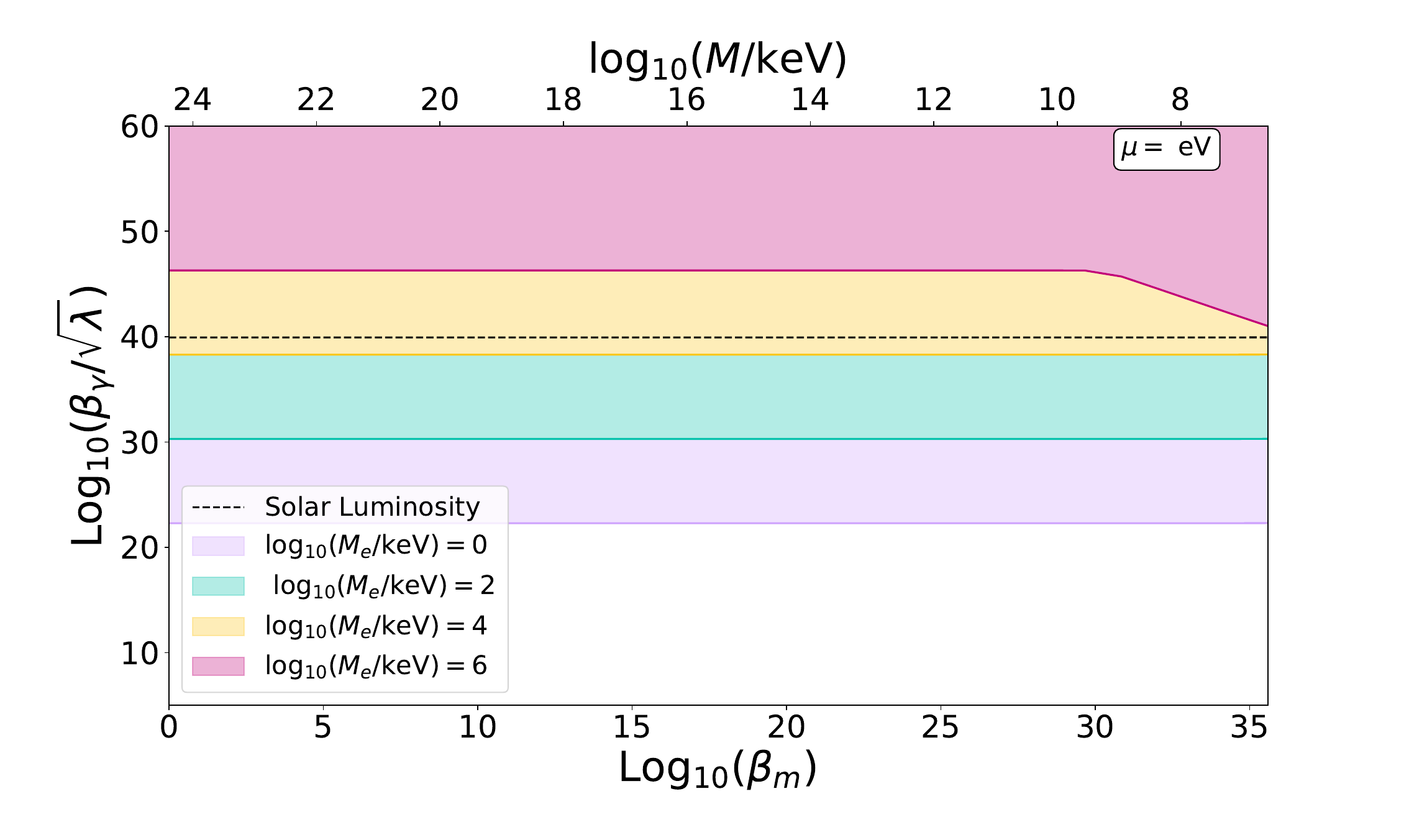}
        \label{fig:eV}
    }
    \subfigure[\,$\mu = 1$\,keV]{
        \includegraphics[width=0.45\linewidth]{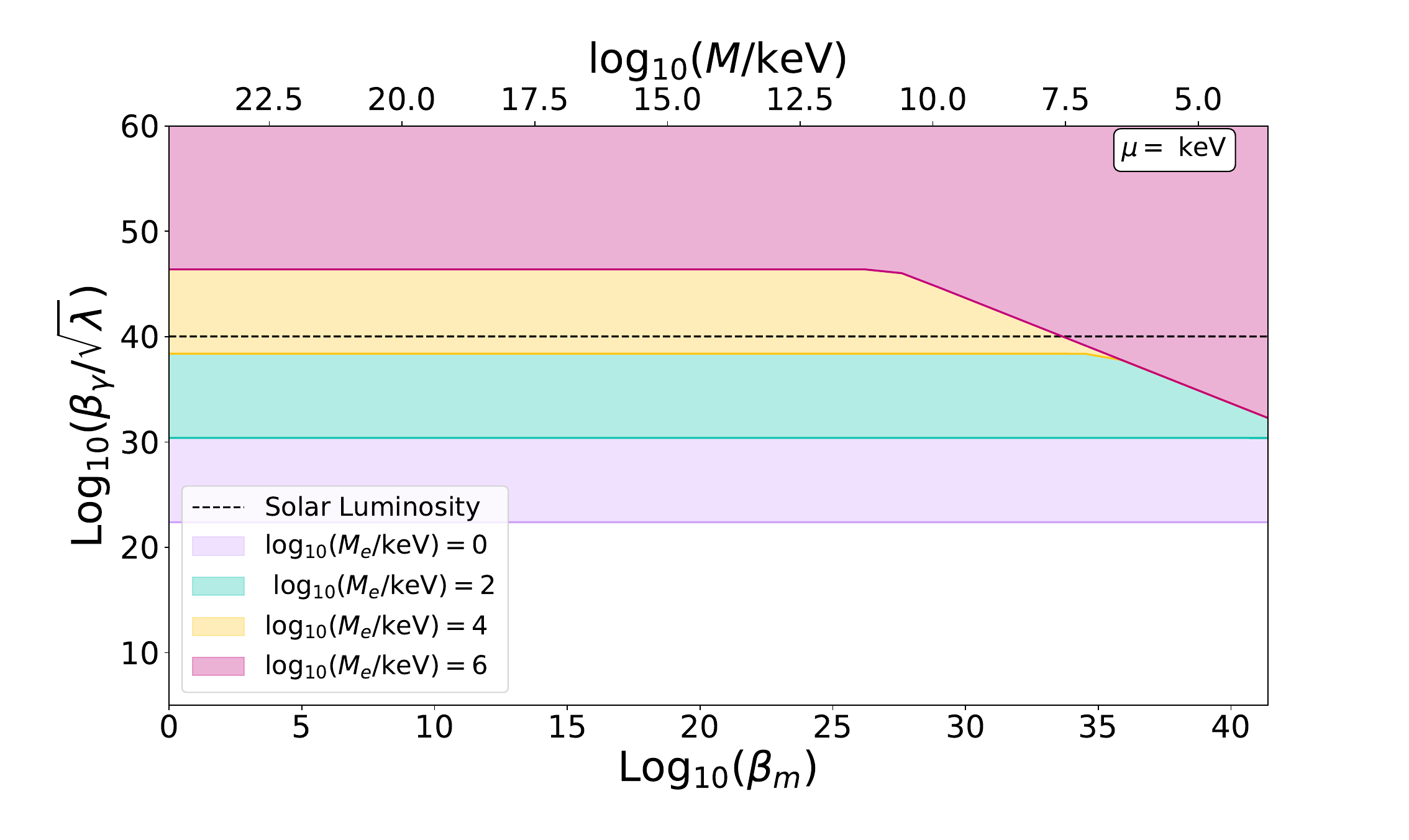}
        \label{fig:keV}
    }
    \subfigure[\,$\mu = 1$\,MeV]{
        \includegraphics[width=0.45\linewidth]{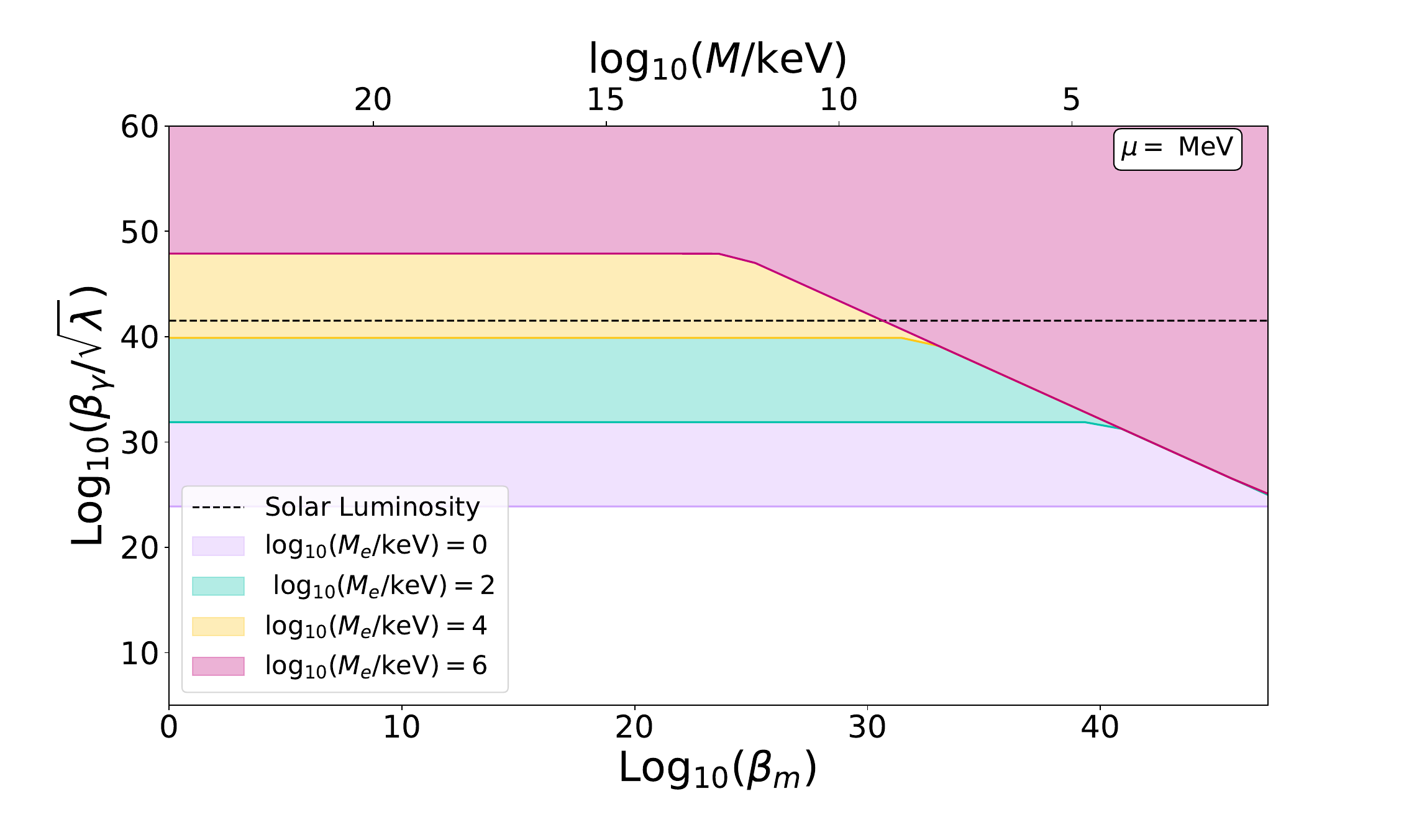}
        \label{fig:MeV}
    }
    \subfigure[\,$\mu = 1$\,GeV]{
        \includegraphics[width=0.45\linewidth]{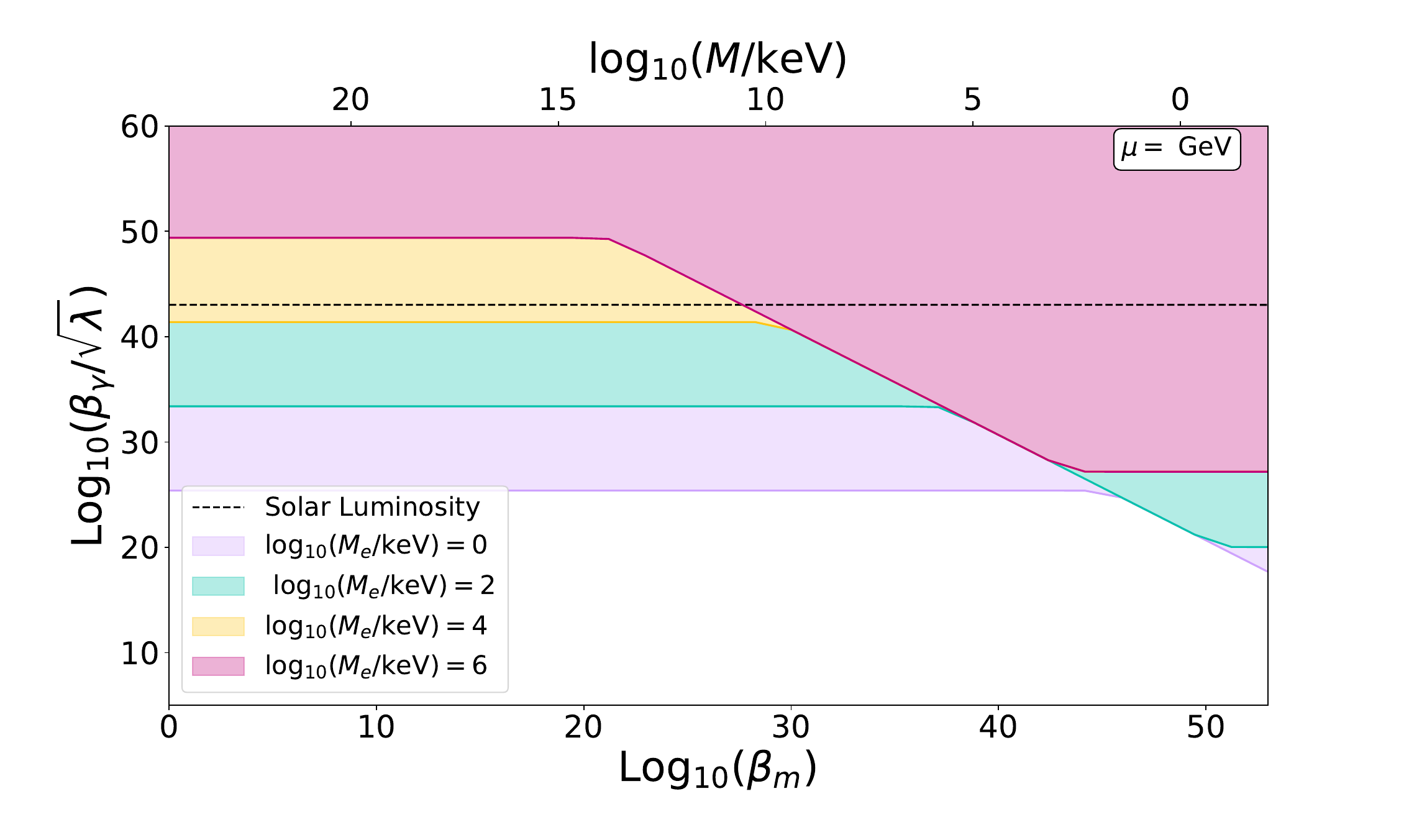}
        \label{fig:GeV}
    }
    \caption{Direct detection bounds on the symmetron parameter space derived from the XENONnT binned data. Each panel shows the allowed region in the $\beta_m\textrm{--}\beta_\gamma/\sqrt{\lambda}$ plane for a fixed value of $\mu$. Solid colored curves correspond to different disformal scales, $\log_{10}(M_e/\mathrm{keV}) = 0,2,4,6$, while the dashed black line indicates the 3\% solar luminosity constraint. The panels correspond to increasing values of $\mu$ from $0.1$\,meV to 1\,GeV. Note that the horizontal scale is different in each panel, to reflect the fact that increasing $\mu$ increases the maximum $\beta_m$ which can be probed with solar production following Eq.~\eqref{eq:con}.}
    \label{fig:directconstraints}
\end{figure*}

Our results are shown in Fig.~\ref{fig:directconstraints}. Each panel displays the direct detection constraints on the $\beta_m\textrm{--}(\beta_\gamma/\sqrt{\lambda})$ plane for a different value of the fundamental scale~$\mu$ ranging from $0.1$\,meV to 1\,GeV. The solid curves in each panel correspond different choices for the disformal scale $\log_{10}(M_e/\mathrm{keV}) \in \{0,2,4,6\}$, with the shaded parameter space above each curve (i.e. at greater $\beta_\gamma/\sqrt{\lambda})$  excluded.  For comparison we show the 3\% solar luminosity constraint for the $\mu$ in question as a dashed black line. The upper horizontal axis on each plot labels the energy scale $M \equiv M_{\rm Pl}/\sqrt{\beta_m}$, which is an alternative parameterization of the symmetron coupling strength to SM matter fields that is often used to report bounds from table-top experiments~\cite{Brax:2022olf, Brax:2026cmh}.

Increasing $\mu$ modifies both the symmetron VEV and the critical density $\rho_*$ in Eq.~\eqref{eq:criticaldensity}, changing the region of the Sun in which the symmetry is broken and production can occur. In particular, we note that as $\mu$ is increased, the maximum value of $\beta_m$ for which  production in the tachocline remains viable, as delineated in Eq.~\eqref{eq:con}, also increases. This is reflected through the differing horizontal scale in each panel. 

The relative importance of the conformal and disformal channels in symmetron absorption in xenon is set by the relative size  of the disformal coupling, controlled by the inverse of the scale $M_e$, compared to the effective conformal coupling which is proportional to $\beta_m \phi_0$. In general, at low values of $\beta_m$, the disformal channel dominates, resulting in a bound that is independent of $\beta_m$. Conversely, at high $\beta_m$ the conformal channel can become dominant. In this regime the reach to $\beta_\gamma / \sqrt{\lambda}$ is no longer constant in $\beta_m$ but improves as this parameter is increased. The value of $\beta_m$ at which the conformal interaction becomes dominant (for any given disformal scale $M_e$) depends on the value of $\mu$, which sets the scale of $\phi_0$ and thus impacts the effective conformal coupling. It thus follows that at smaller $\mu$ the cross-over between the disformal and conformal channels is necessarily pushed to higher $\beta_m$ in order to maintain the same effective conformal coupling. Combined with the fact that the maximum value of $\beta_m$ that can be probed with solar production is smaller at low $\mu$, it follows that for the smallest values of $\mu$ considered here - namely 0.1 meV and  0.1 eV - absorption is dominated by the disformal channel across all viable $\beta_m$, and the conformally dominated regime is never reached. The role that the conformal channel plays in carving out the direct detection landscape becomes significantly more prominent as the scale $\mu$ is increased. As expected, we note that the direct detection reach at low $\beta_m$ (i.e. in the disformally dominated regime)  depends sensitively on the parameter $M_e$, with improved sensitivity at low values of this scale (where the disformal coupling is greatest).  For $M_e \sim\,$GeV we note that the solar luminosity requirement is more constraining than the direct detection limit (at small $\beta_m$, $\mu$), whereas for the other values of $M_e$  considered here it is weaker.


\section{Discussion}
\label{sec:discussion}

In addition to the solar and xenon-based limits presented here, symmetrons have also been tested in controlled laboratory settings. In particular, $\beta_m$ has been constrained through torsion balance experiments~\cite{Adelberger:2006dh, Sabulsky:2018jma}, which seek deviations from the gravitational inverse square law at short distances. Precision atom interferometry experiments~\cite{Brax:2011aw, Burrage:2014oza, Fischer:2024eic,Banks:2025vvz}, Casimir-force measurements~\cite{Brax:2007vm}, and neutron quantum bouncer setups~\cite{Jenke:2014yel} also provide direct sensitivity to short-distance symmetron-mediated forces. These searches are especially powerful in regimes where fifth forces mediated by symmetron exchange are unscreened in the laboratory environment. In such regimes, laboratory experiments can outperform solar-system and astrophysical probes, despite the latter involving considerably heavier source masses, since the symmetron-mediated fifth force may be screened in dense astrophysical environments whilst remaining unscreened in the lab due to carefully engineered conditions. 

Current laboratory limits constrain the couplings $\lambda$ and  $\beta_m$ over a wide range of $\mu$, with future technological advances, such as next-generation atom interferometers~\cite{Banks:2025vvz} and upgraded torsion-balance experiments~\cite{Zhao:2022hwa, Yin:2025uzf}, projected to further extend coverage of this parameter space. However, unlike the solar probes considered in this work, these experiments are not sensitive to $\beta_{\gamma}$, thus underscoring the need for a diverse search program to fully probe symmetron phenomenology.

Taken together, solar, direct detection, and table-top experiments provide a complementary network of constraints on screened scalar fields. The Sun provides a high-intensity source of on-shell symmetrons, direct detection experiments serve as efficient absorption targets for this flux, and precision laboratory setups probe static symmetron-mediated forces in the non-relativistic regime. Whereas the latter are insensitive to the symmetron-photon coupling, both  solar luminosity arguments and the direct detection of the solar symmetron flux, explicitly probe this  interaction, rendering our approach  highly complementary to traditional studies. Since the symmetron-photon coupling enters the relevant observables via the combination $\beta_{\gamma}/\sqrt{\lambda}$, it is not possible, unless one specifies to a particular $\beta_{\gamma}$, to directly compare our exclusion regions (which are naturally expressed in the ($\beta_m$,$\beta_{\gamma}/\sqrt{\lambda}$) plane)  to those from table-top force experiments which instead typically report limits on the space of  $(\beta_m, \lambda) $ couplings~\cite{Brax:2022olf, Brax:2026cmh}. Given the non-trivial interplay between different parameters of the model, a consistent interpretation of symmetron parameter space requires a  joint assessment of solar and laboratory constraints  by way of a multi-parameter global fit. Such a study, which falls beyond the scope of this work, would allow for limits on $\lambda$ and $\beta_{\gamma}/\sqrt{\lambda}$ to be recast into the ($\beta_\gamma$,$\beta_{m}$) plane and a combined exclusion region obtained.

Since screening produces qualitatively different phenomenology across environments, a diverse search program exploiting the rich variety of potential couplings of the symmetron to the SM is essential to maximize coverage of symmetron parameter space.  As we have demonstrated, even in regimes in which conformal couplings are suppressed by small field expectation values, disformal channels remain unaffected, rendering experiments able to probe these latter interactions, including underground direct detection experiments, particularly powerful. It thus follows that  a multi-pronged experimental strategy leveraging the complementarity and non-trivial interplay between different symmetron interactions can robustly probe symmetron parameter space, despite this model being subject to environmental screening. Symmetrons therefore emerge as a promising target for both current and near-future experiments, and merit inclusion in the broader search for light scalar fields at the interface of particle physics, cosmology, and astrophysics.

Although not considered explicitly in this work, for completeness we note that helioscope experiments present a further opportunity to probe solar symmetrons. Whilst no such study has ever been performed for the symmetron, these experiments constitute a well-established channel for probing light scalar particles.  In particular the CAST experiment has been deployed to search for chameleons (assuming that they are only produced in the tachocline) via their conversion into X-rays in the presence of the strong transverse magnetic field in the CAST detector~\cite{Brax:2011wp}. The most recent analysis reports an upper bound on the chameleon-photon coupling of $\beta_\gamma^{\rm (cham)} \lesssim 5.7\times 10^{10}$ at the 95\% confidence level (CL)~\cite{CAST:2018bce}, for chameleon–matter couplings in the range $1<\beta_m^{\rm (cham)}<10^6$. Whilst performing an analogous study for the symmetron demands a detailed simulation of the CAST detector and thus falls beyond the scope of this work, we may obtain a crude estimation of the potential reach of such a study by exploiting the correspondence between the interactions which govern the relevant processes in the symmetron and chameleon theories. 
Concretely, we note that the Lagrangian describing the chameleon-photon interaction which is responsible for both solar production and helioscope detection is 
\begin{equation}
    \mathcal{L}_{\rm cham-\gamma} = \frac{\beta_\gamma^{\rm (cham)}}{M_{\rm Pl}}\,\phi\,F_{\mu\nu}F^{\mu\nu}\,.
\end{equation}
Up to the identification $\beta_\gamma^{\rm (cham)} = \beta_\gamma \phi_0/M_{\rm Pl}$, this is of precisely the same form as the dominant symmetron-photon interaction given in Eq.~\eqref{eq:LagrangianEM2}.  As a result,  solar production and helioscope detection of symmetrons thus proceed analogously to the chameleon case, albeit with an effective coupling constant which is suppressed by the factor $\phi_0/M_{\rm Pl}$. Using this mapping, it is possible to estimate the reach of CAST in symmetron parameter space to be 
\begin{equation}
    \beta_\gamma/\sqrt{\lambda} \lesssim 10^{38}\,\left(\frac{\rm eV}{\mu}\right)\,.
\end{equation}
 Whilst at face value, these estimations appear to be marginally more stringent than the  solar luminosity constraints obtained in this work, it is important to take into account that the CAST chameleon bound was derived under a different set of assumptions.  Whilst the tachocline was similarly taken to be  located at $0.7\,R_\odot$ with width $0.01\,R_\odot$, the magnetic field was instead assumed to be 10 T compared to our 30 T, which, as illustrated in Fig.~\ref{fig:solar constraint} would be expected to yield an even weaker solar luminosity bound. More significantly however, the CAST analysis assumes that the scalar flux emitted from the Sun i.e. their search target, comprises 10~\% of the solar luminosity. As part of their study they compare the observed upper limit on the coupling obtained from a null result in the CAST detector to that from assuming that chameleons cannot account for more than 10 \% of the solar luminosity, finding that at small magnetic field values, including 10 T, the CAST bound is more constraining. At higher magnetic field values however, including at 30 T, they instead find the 10 \% solar luminosity constraint to be a more powerful probe.  One might also expect this conclusion to hold if the luminosity threshold and flux reaching the cast detector were commonly set to 3 \%. Moreover, there are many further subtleties that this simple rescaling argument does not capture. For instance both the solar production and helioscope conversion rates depend on the effective mass of the scalar field, which parametrically differs between the two models. In the case of the symmetron, $m_s$ is governed by $\mu$, introducing an additional, non-trivial dependence of the relevant rates on this parameter which is not accounted for by simply rescaling the coupling. Nonetheless this simple estimate serves to show that bounds from helioscope experiments could offer similar constraining power to symmetron models as limits derived from luminosity arguments and thus warrant a dedicated study.

 Whilst this work forms an important first step towards fully exploiting the Sun as a testable symmetron source, several refinements are in order. In particular we emphasize that our study only considers production in the tachocline, however for much of the parameter space we consider, production is expected to continue further into the core.  In the case of chameleons, it was shown that this additional production, which, due to the higher densities involved can also occur via Primakoff-like processes mediated by the microscopic magnetic fields of charged ions in the solar plasma, can become important ~\cite{OShea:2024jjw}, and an extended analysis is required to investigate the extent to which this carries over to the symmetron. For this reason, our bounds should be viewed as a conservative first estimate, which may become stronger should a more comprehensive treatment accounting for production further into the solar interior, as performed in Ref.~\cite{Yuan:2025twx} for the chameleon, be undertaken. Whilst a back-of-the-envelope estimate indicates that a dedicated application of the CAST data to the symmetron could be fruitful, we also note that the forthcoming BabyIAXO helioscope experiment will soon begin data taking. A future combined study incorporating both of these is expected to further improve the sensitivity to symmetron models.


Before concluding, it is important to assess from a theoretical perspective whether the symmetron potential is radiatively stable or if loop corrections modify the mass or screening behavior in the environments relevant for our analysis. As we have seen, expanding the symmetron Lagrangian around the VEV generates a coupling to electrons of the form $y_e \sim \beta_m m_e\phi_0/M_{\rm Pl}^2$, which induces both thermal and vacuum corrections to the scalar mass. The thermal correction can be estimated to be~\cite{Babu:2019iml}
\begin{equation}
\begin{split}
    \Delta m_{s,{\rm th}}^2 &= y_e^2\,\frac{n_e}{m_e} \\
    &\approx (13{\rm\,eV})^2\left(\frac{\beta_m/\sqrt{\lambda}}{10^{50}}\right)^2\left(\frac{\mu}{\rm eV}\right)^2\left(\frac{n_e}{10^{23}\,\rm cm^{-3}}\right)\,,
\end{split}
\end{equation}
which captures the leading effect of the electron bath on the symmetron mass. The corresponding vacuum Coleman–Weinberg correction takes the standard form
\begin{equation}
\begin{split}
    \Delta m_{s,{\rm vac}}^2 &= -\frac{y_e^2m_e^2}{4\pi^2}\left(\ln\frac{m_e^2}{\mu^2}-1\right)\\
    &\approx -\frac{\beta_m^2 m_e^4\mu^2}{4\pi^2M_{\rm Pl}^4\lambda}\left(\ln\frac{m_e^2}{\mu^2}-1\right)\,.
\end{split}
\end{equation}
In addition, scalar self-interactions generate one-loop corrections through the Coleman–Weinberg mechanism of spontaneous symmetry breaking~\cite{Burrage:2016xzz}. For the parameter space considered here, each of these contributions is expected to remain subdominant. However, for symmetron models with extremely large values of $\beta_m/\sqrt{\lambda}$, the loop-induced corrections can become comparable to the tree-level mass term $\sim \mu^2$ when
\begin{equation}
    \frac{\beta_m}{\sqrt{\lambda}} \gtrsim 10^{43}\,\min\left\{1, \left(\frac{n_e}{\rm 10^{31}\,cm^{-3}}\right)^{-1/2}\right\}\,,
\end{equation}
potentially modifying the effective mass and screening behavior. In this regime, an appropriate tuning of the bare parameters would be required to realize screening. 

\section{Conclusions}
\label{sec:conclusions}

In this work, we have investigated the solar production of symmetrons and derived new bounds on their parameter space using both solar luminosity considerations and direct detection data. We showed that photon-symmetron conversion in the solar tachocline generates a sizable flux of keV-scale symmetrons reaching Earth. Demanding the associated energy loss to remain below a conservative 3\% of the total solar luminosity  yields robust upper limits on the conformal symmetron-photon coupling across a broad range of fundamental scales. We further demonstrated that the solar symmetron flux can be absorbed in liquid xenon detectors, resulting in detectable electronic recoils. Using the binned data from XENONnT as a benchmark for xenon-based experiments, we obtained complementary bounds on symmetron interactions, stressing the distinct roles of conformal and disformal couplings. Whilst solar emission mainly constrains the conformal symmetron-photon coupling, both conformal and disformal symmetron-electron interactions can play a role in xenon absorption. The disformal operator dominates both at high recoil energies, and in regions of parameter space in which conformal interactions are suppressed by small field expectation values. Our work substantially tightens the viable symmetron parameter space, with further gains expected should a more comprehensive treatment including production further into the solar interior be undertaken in the future.

Our analysis emphasizes the complementarity of astrophysical, underground, and laboratory probes. Solar emission provides a high-intensity natural source of symmetrons, xenon detectors offer sensitivity to the absorption of this flux, and precision table-top experiments constrain symmetron-mediated forces in controlled terrestrial settings. Together, these methods establish a coherent and expanding program to search for environmentally screened scalar fields. The results presented here constitute both the first dedicated investigation of solar symmetron production and the first set of constraints on the symmetron–photon coupling to date, opening the window to large regions of unexplored symmetron parameter space. Future work could refine this analysis to include improved solar magnetic field models, particularly the magnetic field profile and plasma properties, along with refined stellar structure inputs and updated direct detection data. Progress in experimental sensitivity will increasingly constrain the parameter space of symmetron models, and screened light scalars more broadly, allowing for a more incisive exploration of the physics of the dark sector.


\vspace{.3cm}
\begin{acknowledgments}
We thank Clare Burrage and Sunny Vagnozzi for fruitful discussions throughout all stages of this work. This publication is based upon work from the COST Actions ``COSMIC WISPers'' (CA21106) and ``Addressing observational tensions in cosmology with systematics and fundamental physics (CosmoVerse)'' (CA21136), both supported by COST (European Cooperation in Science and Technology). H.B.\ and A.C.D.\ acknowledge partial support from the Science and Technology Facilities Council (STFC) through STFC consolidated grant ST/T000694/1. L.V.\ acknowledges support by Istituto Nazionale di Fisica Nucleare (INFN) through the Commissione Scientifica Nazionale 4 (CSN4) Iniziativa Specifica ``Quantum Universe'' (QGSKY), the National Natural Science Foundation of China (NSFC) through grant no.\ 12350610240 ``Astrophysical Axion Laboratories'', and the State Key Laboratory of Dark Matter Physics at Shanghai Jiao Tong University. L.V.\ additionally thanks the Tsung-Dao Lee Institute and the Xplorer Symposia Organization Committee of the New Cornerstone Science Foundation for hospitality during the final stages of this work. This work made use of the open source software matplotlib~\cite{2007CSE.....9...90H}, numpy~\cite{2020Natur.585..357H}, and scipy~\cite{2020NatMe..17..261V}.
\end{acknowledgments}

\appendix

\section{Derivation of the production rate}
\label{sec:appendix}

The calculation of the production rate proceeds via the finite--temperature field theory relation between the imaginary part of the scalar self--energy $\Pi_\phi$ and the emission rate. The photon propagating in the solar plasma is described by the polarization tensor
\begin{equation}
    \Pi^{\mu\nu}(\omega,\mathbf{k}) = \Pi_T(\omega,k) P_T^{\mu\nu} + \Pi_L(\omega,k) P_L^{\mu\nu}\,,
\end{equation}
where $P_{T,L}^{\mu\nu}$ are the usual transverse and longitudinal projectors with respect to the plasma rest frame. The poles of the corresponding propagator define the dispersion relations
\begin{equation}
    \omega^2 - k^2 - \Re\,\Pi_{T,L}(\omega,k) = 0\,,
\end{equation}
corresponding to transverse and longitudinal (plasmon) modes. The photon spectral functions entering the thermal calculation are given by
\begin{equation}
    \label{eq:rhoTL}
    \rho_{T,L}(\omega,k) = \frac{-2\,\Im\,\Pi_{T,L}(\omega,k)}{[\omega^2-k^2-\Re\,\Pi_{T,L}(\omega,k)]^2 + [\Im\,\Pi_{T,L}(\omega,k)]^2}\,.
\end{equation}
The imaginary part $\Im\,\Pi_{T,L}$ encodes both pole (plasmon) and continuum (Landau damping) contributions. 

Following Weldon’s theorem~\cite{Weldon:1983jn}, the rate per unit phase space for producing symmetron fluctuations $\delta\phi$ of energy $\omega$ is
\begin{equation}
    \label{eq:weldon}
    \Gamma^{\rm prod}_\phi(\omega,\mathbf{k}) = -\frac{\Im \Pi_\phi(\omega,\mathbf{k})}{\omega}\,f_B(\omega)\,,
\end{equation}
where $f_B(\omega)=1/(e^{\omega/T}-1)$ is the Bose-Einstein distribution. At leading order, the scalar self-energy $\Pi_\phi$ arises from the mixing of the scalar with a photon in the presence of two insertions of the external magnetic field $\mathbf{B}$. Only the transverse photon polarization contributes at this order. Explicitly, the imaginary part of the scalar self-energy is
\begin{equation}
    \Im\,\Pi_\phi(\omega,\mathbf{k}) = \left(\frac{\beta_\gamma\phi_0}{M_{\rm Pl}^2}\right)^2 B_\perp^2 \, \omega^2 \, \rho_T(\omega,k)\,,
\end{equation}
where $B_\perp$ is the component of the magnetic field transverse to the direction of propagation. The factor $\omega^2\rho_T$ originates from the photon propagator contracted with the external field insertions.

\begin{figure}[thb]
\centering
\begin{tikzpicture}
    \begin{feynman}
        \vertex (a) {\(\phi\)};
        \vertex [right=of a] (b);
        \vertex [right=of b] (c);
        \vertex [right=of c] (d) {\(\phi\)};
        \vertex [below=of b, crossed dot] (B1) {\(\)};
        \vertex [below=of c, crossed dot] (B2) {\(\)};

        \diagram* {
            (a) -- [scalar] (b) -- [photon, edge label = \(\gamma\)] (c) -- [scalar] (d),
            (b) -- [photon, edge label = \(\mathbf{B}\)] (B1),
            (c) -- [photon, edge label = \(\mathbf{B}\)] (B2),
        };
    \end{feynman}
\end{tikzpicture}
\caption{Lowest--order contribution to the symmetron self--energy in a plasma in the presence of an external magnetic field. The crossed vertices denote insertions of the background field $\mathbf{B}$. Each scalar--photon vertex carries the effective coupling $\beta_\gamma \phi_0 / M_{\rm Pl}^2$arising from the expansion around the background field value.}
\label{fig:feynman_TFT}
\end{figure}
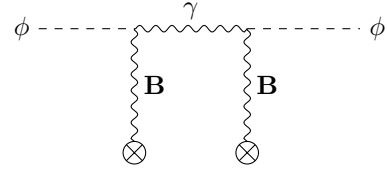

Approximating the transverse photon damping rate in the quasi-particle approximation for the transverse photon mode,
\begin{equation}
    \Im\,\Pi_T(\omega,k)\simeq -\omega\,\Gamma_\gamma(\omega,r)\,,    
\end{equation}
and inserting Eq.~\eqref{eq:rhoTL} into the Weldon relation Eq.~\eqref{eq:weldon} gives the symmetron production rate per mode in the weak-coupling limit,
\begin{equation}
    \Gamma^{\rm prod}_\phi(\omega,k)
    \simeq \frac{2\left(\frac{\beta_\gamma\phi_0B_\perp}{M_{\rm Pl}^2}\right)^2\,\omega^2\,\Gamma_\gamma(\omega,r)\,f_B(\omega)}
    {(\omega^2-k^2-\Re\,\Pi_T)^2+(\omega\,\Gamma_\gamma(\omega,r))^2}\,,
\end{equation}
where the sum over transverse photon polarizations is implicitly included. Near resonance, this reduces to the familiar Breit–Wigner form. Resonant enhancement occurs when the effective photon mass matches the symmetron mass, $m_\gamma \simeq m_s$, leading to a peak in the production rate. The damping rate $\Gamma_\gamma$ encodes photon production and absorption processes in the solar medium. Following Ref.~\cite{OShea:2024jjw}, we retain only Thomson scattering and free-free (bremsstrahlung) contributions, which dominate in the keV range. The differential emission rate is obtained by integrating over phase space and imposing the on-shell relation $\omega^2 = k^2 + m_s^2$ to yield
\begin{equation}
    \frac{\mathrm{d}\dot N}{\mathrm{d}\omega} \!=\! \frac{2 \beta_\gamma^2\phi_0^2}{\pi M_\mathrm{Pl}^4} \!\int_{R_-}^{R_+}\!\!\mathrm{d}r r^2 B_\perp^2(r) \frac{\omega(\omega^2 - m_s^2)^{3/2}}{(m_\gamma^2 - m_s^2)^2 \!+\! (\omega \Gamma_\gamma)^2} \frac{\Gamma_\gamma}{e^{\omega/T} - 1}\,.
    \label{eq:dN_TFT--2}
\end{equation}
In this expression we have set $\Re\,\Pi_T\simeq m_\gamma^2(r)$, the local plasma frequency squared, and explicitly displayed the photon damping/absorption rate $\Gamma_\gamma(\omega,r)$ which arises from $\Im\,\Pi_T$. Integrating Eq.~\eqref{eq:dN_TFT--2} over the tachocline volume, assumed to be a thin shell, leads to the expression in Eq.~\eqref{eq:dN_TFT2} in the main text. In Fig.~\ref{fig:feynman_TFT} we show the lowest-order contribution to the symmetron self-energy in a magnetized plasma. The diagram represents the mixing of the scalar field with a transverse photon in the presence of an external magnetic field, mediated by the effective symmetron-photon interaction proportional to $\phi_0/M_{\rm Pl}^2$.

\bibliographystyle{apsrev4-1}
\bibliography{refs.bib}

\end{document}